\documentclass[3p,preview,authoryear]{elsarticle}
 
\usepackage{booktabs}           
\usepackage{tikz}               
\usepackage{mathtools}               

\usepackage{lineno,hyperref}
\usepackage[symbol]{footmisc}

\usepackage{sectsty}
\usepackage{fullpage}
\usepackage{graphicx}
\usepackage{caption}
\usepackage{epstopdf}
\usepackage{epsfig}
\usepackage{pdfpages}
\usepackage{float}
\usepackage{amsmath,bm}
\usepackage{textcomp}
\usepackage{upgreek}
\usepackage{multirow}
\usepackage{textcomp}
\usepackage{accents}
\usepackage{amsfonts}
\usepackage{lscape}
\usepackage{comment}

\modulolinenumbers[1]

\journal{arXiv}

\begin{document}

\begin{frontmatter}

\title{Full-field, quasi-3D hygroscopic characterization of paper inter-fiber bonds}

\author[1]{N.H. Vonk} \author[1]{R.H.J. Peerlings} \author[1]{M.G.D. Geers} \author[1]{J.P.M. Hoefnagels*} 

\cortext[mycorrespondingauthor]{J.P.M. Hoefnagels}
\ead{J.P.M.Hoefnagels@tue.nl}

\address[1]{Department of Mechanical Engineering, Eindhoven University of Technology, Eindhoven, The Netherlands}

\begin{abstract}
The state-of-the-art in paper mechanics calls for novel experimental data covering the full-field hygro-expansion of inter-fiber bonds in paper, i.e., the 3D morphological changes and inter-fiber interactions. Therefore, a recently developed full-field single fiber hygro-expansion measurement methodology based on global digital height correlation is extended to orthogonally bonded inter-fiber bonds, to investigate their full-field quasi-3D hygroscopic behavior. A sample holder has been developed which enables the quasi-3D characterization of the initial geometry of individual inter-fiber bonds, including the fiber thickness and width along the length of the fibers as well as the degree of wrap around and contact area of the bond, which are vital for understanding the inter-fiber bond hygro-mechanics. Full-field hygroscopic testing revealed novel insights on the inter-fiber interactions: (i) the transverse hygro-expansion of each fiber strongly reduces when approaching the bonded area, due to the significantly lower longitudinal hygro-expansion of the other bonded fiber. (ii) The relatively large transverse strain of one fiber stretches the other crossing fiber in its longitudinal direction, thereby significantly contributing to the sheet scale hygro-expansion. (iii) Out-of-plane bending is observed in the bonded region which is driven by the significant difference in transverse and longitudinal hygro-expansion of, respectively, the top and bottom fiber constituting the bond. A bi-layer laminate model is derived to rationalize the bending deformation and an adequate match is found with the experimental data. Under the assumption of zero bending, which represents constrained inter-fiber bonds inside a paper sheet, the model is able to predict the contribution of the transverse strain in the bonded regions to the sheet-scale hygro-expansion. 
\end{abstract}


\end{frontmatter}

\newpage
\section{Introduction}
The moisture-induced dimensional changes, i.e. the hygro-expansion, of paper-based products has been studied extensively over the past decades, but still leaves many open questions for many applications \citep{alava2006physics, fellers2007interaction}. For instance, better understanding of the hygro-expansion of paper is essential for printing applications to minimize unwanted macroscopic out-of-plane deformations of the paper sheet known as fluting, cockling, and curling \citep{kulachenko2005tension, bosco2018role, kulachenko2021moisture}. These unwanted moisture-driven phenomena are governed by a cascade of complex mechanisms occurring at all relevant underlying length scales, i.e. at the scale of single fibers, inter-fiber bonds, and fiber networks.\\ \indent
Because testing the microscopic paper components is challenging, most literature focuses on in-plane sheet-scale hygro-expansion measurements, complimented by inter-fiber bond models to explain the findings, see, e.g., \citep{fellers2007interaction, uesakaQi1994hygroexpansivity, niskanen1997dynamic, larsson2008influence}. These sheet-scale hygro-expansion tests typically involve measuring the dimensions as a function of the relative humidity (RH) \citep{fellers2007interaction}. The micro-scale natural fibers are the main constituent of the paper sheet, and their hygro-expansion, which is significantly larger (20$-$40 times) in transverse compared to longitudinal direction \citep{fellers2007interaction, vonk2021full}, drives the hygro-expansion of the paper sheet \citep{page1962new}. Back in 1994, \cite{uesaka1994general} stated that, for machine paper (containing strongly aligned fibers in the machine direction (MD)), the longitudinal hygro-expansion of the fibers is comparable to the sheet-scale hygro-expansion in MD, indicating that the longitudinal fiber hygro-expansion governs the sheet scale in MD. In contrast, for non-oriented handsheets, yielding isotropic sheet-scale hygro-expansion, the longitudinal hygro-expansion of the fibers mainly contributes in the freestanding segments, whereas the significantly larger transverse hygro-expansion is transferred in the bonded regions in order to contribute to the sheet \citep{wahlstrom2009development, berglund201112}. This bond strain transferal was experimentally studied first by \cite{page1962formation} and later by \cite{nanko1995mechanisms}. The authors demonstrated experimental evidence of the transferal of the relatively larger transverse strain of one fiber to another fiber in the bonded regions (referred to as transverse strain transferal). However, the results were not entirely conclusive regarding quantification of the transverse strain transferal, which was roughly estimated to be up to 50\%.\\ \indent
A modeling approach also allows describing the inter-fiber bonds and investigate the transverse strain transferal. To this end, multiple fiber network models describing the moisture-induced deformations have been proposed over the years, e.g., idealized homogenization network models \citep{bosco2015explaining, bosco2015predicting}, random (homogenization) network models \citep{strombro2008mechano, bosco2016local, bosco2017asymptotic}, 2D level set XFEM network model \citep{samantray2020level}, and 3D beam network models \citep{sellen2014mechanical, motamedian2019simulating, brandberg2020role}. For instance \cite{brandberg2020role} proposed a 3D beam network model to distinguish the contribution of the longitudinal and transverse fiber hygro-expansion coefficient to the sheet-scale hygro-expansion. It was demonstrated that the longitudinal hygro-expansion coefficient (chosen as $\beta_l$ = 0.03) of the fibers was directly taken up by the network, whereas only \textsuperscript{$\sim$}4\% of the transverse hygro-expansion coefficient ($\beta_t$ = 0.6) was transferred, resulting in an isotropic sheet-scale hygro-expansion coefficient of ($\beta_l$ + 0.04$\beta_t \approx$) 0.054, directly emphasizing the relatively weak contribution of the significantly large transverse fiber hygro-expansion to the network. However, while all these models have their own merits, they all require experimental identification and validation, in particular for the most significant and difficult aspect to characterize the hygro-expansion behavior of inter-fiber bonded regions. \\ \indent
While the bond strength of isolated inter-fiber bonds, i.e., mode 1 and 2 fracture \citep{stratton1990dependence, schmied2012joint, fischer2012testing, schmied2013holds, jajcinovic2016strength, jajcinovic2018influence}, inter-fiber bond formation mechanisms \citep{nanko1989mechanisms, hirn2015comprehensive}, fiber-to-fiber contact surface measurements \citep{gilli2009optical, kappel2009novel, kappel2010revisitingA, kappel2010revisitingB}, and complete 3D inter-fiber bond geometries including contact surface \citep{sormunen2019x} have been characterized with great detail, direct data on the hygroscopic behavior of isolated inter-fiber bonds is missing. As stated before, \cite{page1962formation} were the first to analyze the transverse strain transferal, and stated that transverse shrinkage of a fiber causes increased longitudinal contraction, thus forming micro-compression in the bonded region identified as wrinkles. However, the authors could not quantify the portion of transverse strain which was transferred to the other fiber. \cite{nanko1995mechanisms} followed up on this work and studied the fiber shrinkage in freestanding and bonded segments during paper formation by determining the length difference between silver grains applied to (near orthogonally bonded) fiber networks in the wet (moisture content of 60\%) and dry state (RH = 60\%). The authors reported that the freestanding fiber segments shrunk significantly less compared to the bonded segments, confirming the transverse strain transferal. Identification of the transverse shrinkage of the freestanding fiber segments in the wet webs enabled a quantitative comparison between the bond strains, from which the authors extracted a transverse strain transferal of up to 50\%. This is much larger than the contribution of \textsuperscript{$\sim$}4\% found in the 3D network model proposed by \cite{brandberg2020role}. The large measurement uncertainty in the results of \cite{nanko1995mechanisms} stems from the accuracy of the silver grain method in which relatively small strains are determined over short lengths (i.e., 30$-$100 $\mu$m), containing considerable topographical fluctuations. Moreover, the dry observations were only compared to the fully wet state, where the inter-fiber bonds are weak and not properly formed, entailing significant morphological changes at the bonding surface with fiber-to-fiber in-plane rotations. In contrast, the geometry is relatively "fixed" during hygro-expansion simulations with the 3D network model. Furthermore, \cite{nanko1995mechanisms} characterized the inter-fiber bonds inside the paper sheet, where each bond is influenced by its neighboring fibers, making it difficult to unravel the individual contributions to the network. \\ \indent
Because the work of \cite{nanko1995mechanisms} showed such intriguing and important results for the paper community, in this work we aim to exploit the advances in microscopy and image correlation techniques over the past decades, to device a new methodology to quantitatively characterize the full-field 3D hygro-expansion of isolated inter-fiber bonds during the full wetting and drying cycle. This results in full-field strain maps in and around the bonded area, enabling the validation of the above-described network models. The recently developed full-field methodology based on Global Digital Height Correlation (GDHC), that enables the characterization of the single fiber hygro-expansion with great precision and robustness is instrumental in reaching this goal \citep{vonk2020robust}. By extending this methodology towards single (isolated) inter-fiber bonds, it should be possible to measure the full-field 3D hygro-expansion, including the variation of the strain field around the bonded area, transverse strain transfer in the bonded region, and comparison of freestanding to bonded segment characteristics during wetting and drying. \\ \indent
To achieve this goal, the following key challenges need to be overcome: (i) isolation, clamping and preparation of inter-fiber bonds, which poses a challenge compared to single fibers, (ii) in order to understand the mechanics of the inter-fiber bonds, the initial 3D geometry is required, i.e., thickness of the bonded fibers, degree of wrap around, and an estimation of the bonded area , and (iii) the kinematic regularization used to describe the displacement fields in the GDHC algorithm must be improved, as significantly larger strain gradients are expected than for single fibers. To address challenge (i) and (ii), a custom sample holder has been designed that allows for double sided imaging of isolated orthogonal inter-fiber bonds to obtain a quasi-3D high-resolution reconstruction of the initial geometry. Challenge (iii) is addressed by performing a polynomial order variation to find the correct strain fields \citep{vonk2020robust, neggers2014direct, hoefnagels2022accurate}. It will be shown that this methodology enables the identification of the full-field 3D hygro-expansion of inter-fiber bonds and their mechanics, covering (i) accurate longitudinal and transverse strain gradient around the bonded area and (ii) the magnitude of the transverse strain transferal from fiber-to-fiber in the bonded region. Moreover, a bi-layer laminate model of the bonded area will be proposed to describe the inter-fiber bond mechanics, which is used to validate and improve the (commonly adopted) fiber stiffness parameters used in previous inter-fiber bond models \citep{brandberg2020role, magnusson2013numerical}. The model also allows to rationalize and predict the transverse strain transferal inside the paper sheet. 

\section{Materials and Methods}

\subsection{Preparation of inter-fiber bonds and single fibers}
\label{ssec:prep}
Isolated inter-fiber bonds are formed from softwood pulp following the same method as proposed by \cite{forsstrom2005influence} and \cite{kappel2009novel}. The preparation involves applying small droplets of highly diluted water-pulp mixture in between a sandwich of two Teflon (PTFE) sheets which is subsequently dried using a conventional sheet dryer, resulting in a sparse network of fibers from which single fibers or inter-fiber bonds can be isolated by carefully cutting the connecting fibers. A nominal pressure (20 kPa) is applied, however, the local pressure applied to the fibers and fiber joints in the sparse web, initiating inter-fiber bond formation, is significantly higher. Since the paper sheet hygro-expansivity depends on the amount of stress applied during drying \citep{lindner2018factors}, the differences in local pressure applied to the fiber joints in the sparse networks may lead to a systematic offset in the hygro-expansivity of the fibers. In order to properly interpret the inter-fiber bond hygro-expansion results, it is essential to know the scatter in fiber hygro-expansion prior to investigating the fiber-to-fiber (hygro-expansion) interactions. Hence, the hygro-expansion of single fibers isolated from the created sparse networks are tested first. Note that the pulp consists of the same mixture of spruce and pine as studied in \citep{vonk2021full}.

\subsection{Single fiber hygro-expansion methodology}
\label{ssec:fibermet}
Single fibers are isolated from the pressed sparse networks by means of cutting and subsequently clamped onto a glass slide using two highly compliant 50 $\mu$m nylon wires, following the methodology proposed in \citep{vonk2020robust}. To enable Global Digital Height Correlation (GDHC), a random pattern of (500 nm polystyrene) nano-particles is applied to the fibers using a dedicated mystification setup \citep{shafqat2021cool}. The prepared fibers are tested individually inside a climate chamber underneath an optical profilometer, in which the RH around the fiber is controlled and cyclically varied from 30 to 90\% RH, while consecutive topographies of the fiber top surface are captured to measure the 3D deformation over time. After testing, the obtained topographies are processed using the GDHC algorithm to extract the displacement fields $U_x$, $U_y$ and $U_z$. The GDHC algorithm uses so-called shape functions, in this case 3D spatial polynomial functions, to map the deformed configuration back to the initial. The optimal amplitudes of these shape functions are found using an iterative modified Newton-Raphson scheme in which the so-called "correlation residual", i.e. the difference between the initial and the back-deformed configuration, is minimized \citep{neggers2016image}. By finding the minimal correlation residual, the 3D displacement fields are obtained, which are combined with the initial fiber topography data to enable quantification of the full-field quasi-3D hygro-expansion, i.e. the longitudinal ($\epsilon_{ll}$), transverse ($\epsilon_{tt}$) and shear ($\epsilon_{lt}$) components of the surface strains, i.e. strains that are computed along the curving fiber surface \citep{shafqat2018bulge}. More details on the testing methodology and GDHC algorithm can be found in \citep{neggers2014direct, neggers2016image, vonk2020robust}. \\ \indent
Three fibers are subjected to four RH cycles, each varying from 30$-$90$-$30\%, as given in Figure \ref{fig:strains_fiber}. A linear slope of 0.5\%/min is used to change the RH set-point, which is kept constant for 2 hours to study possible creep effects at high and low RH levels, similarly to the experiments conducted in \citep{vonk2021full}. Regarding acquisition, a magnification 55$\times$ (100$\times$ objective with 0.55$\times$ field-of-view (FOV) multiplier) was used, reducing the FOV to 147$\times$111 $\mu$m\textsuperscript{2} (1400$\times$1040 pixels), which is adequate to capture the nano-particles with sufficient detail and visualize enough fiber surface.

\subsection{Single fiber hygro-expansivity}
\label{ssec:fiberchar}
\begin{figure}[H]
	\centering
	\includegraphics[width=0.5\textwidth]{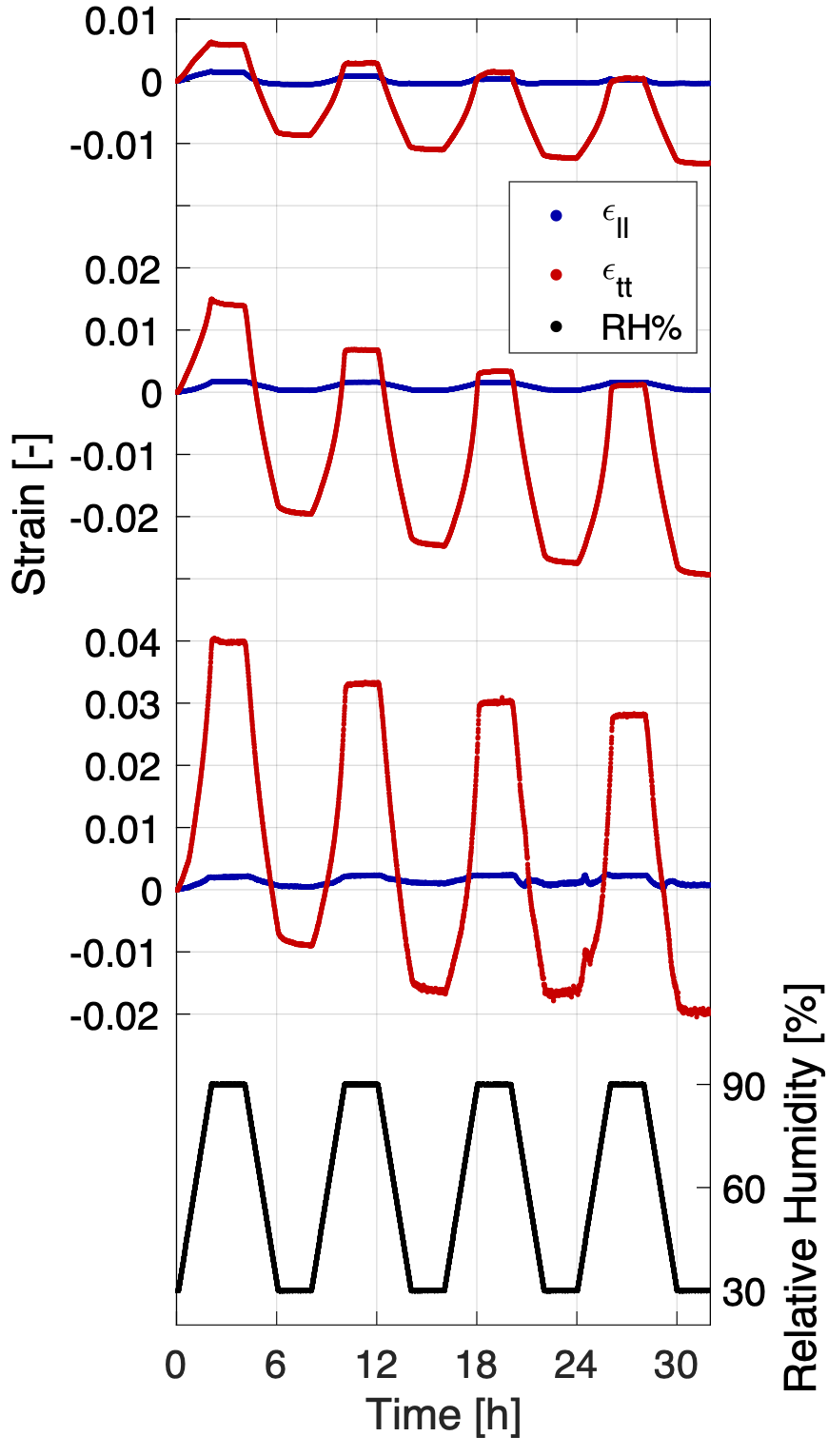}
	\caption{Hygroscopic response of single fibers prepared according to the methodology proposed by \cite{forsstrom2005influence} and \cite{kappel2009novel} subjected to four cycles, each ranging from 30 to 90\% RH. The responses show a large fiber-to-fiber scatter in hygro-expansion which is attributed to the drying stress subjected to the fiber. The overall shrinking curve of the fibers is driven by the release of the dried-in strain stored inside the fiber prior to testing.}
	\label{fig:strains_fiber}
\end{figure}

The hygroscopic responses of the tested fibers are shown in Figure \ref{fig:strains_fiber}. The fiber-to-fiber scatter of the transverse (and longitudinal) hygro-expansivity is remarkably large, i.e. an average strain difference of \textsuperscript{$\sim$}1.5\% for fiber 1 to \textsuperscript{$\sim$}5\% for fiber 3 is found. In \citep{vonk2021full}, it was demonstrated that the transverse hygro-expansion of the five softwood fibers (of the same pulp) picked directly from the pulp bale was 5.8$\pm$0.6\%, demonstrating a significantly lower scatter. Furthermore, the ongoing transverse shrinkage suggests that the three fibers in Figure \ref{fig:strains_fiber} were subjected to a large drying stress, imposing an irreversible initial strain after drying, known as dried-in strain, which is released during the successive wetting and drying cycles. Fiber 3 shows a larger hygro-expansion than fiber 1 for the same RH change, which may be attributed to the amount of (positive) dried-in-strain stored in the fiber structure. It is most likely that the fiber with the highest hygro-expansivity was subjected to the lowest drying stress, because a stress-free fiber has more freedom to swell compared to a fiber subjected to a larger dried-in strain. \\ \indent
These experiments demonstrate that a large scatter in hygro-expansivity is found for fibers which are prepared following the inter-fiber bond methodology proposed by \cite{forsstrom2005influence} and \cite{kappel2009novel}. Hence, it may be expected that this scatter in hygro-expansion will also be present in the fibers of the inter-fiber bonds. 
\noindent

\section{Method extension to inter-fiber bonds}
\subsection{Initial inter-fiber bond characterization}
\begin{figure}[H]
	\centering
	\includegraphics[width=\textwidth]{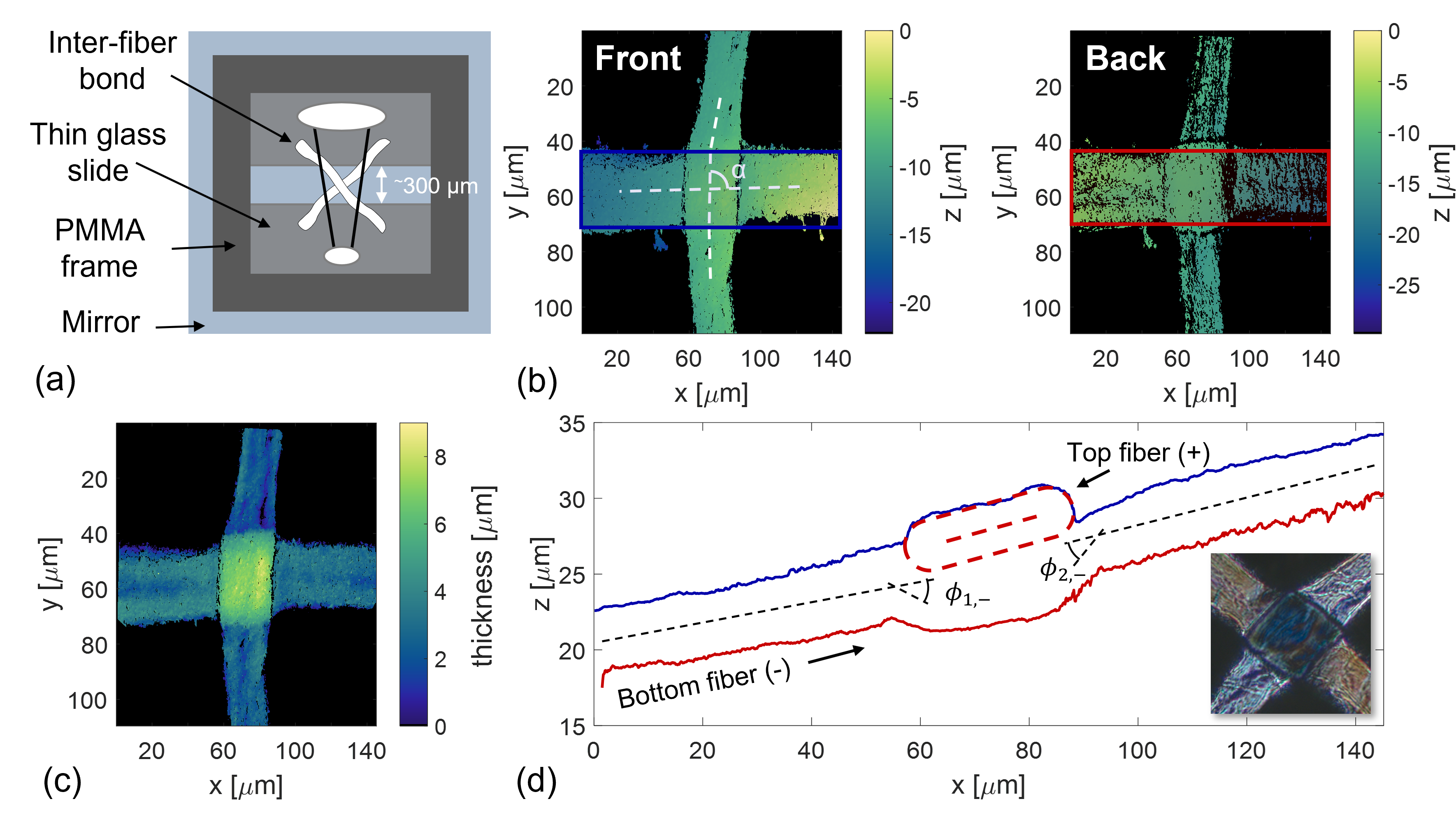}
	\caption{Initial characterization of an inter-fiber bond: (a) schematic of the sample holder which allows double sided imaging of the inter-fiber bond \citep{Maraghechi2023mirror}, (b) front and back topography of an inter-fiber bond with fiber-to-fiber bond angle $\alpha$, (c) thickness map, and (d) fiber width averaged height profile of the front and back of the bottom (horizontal) fiber which allows determination of the interface angle $\phi_{1,-}$ and $\phi_{2,-}$ for the bottom fiber and polarization optical microscopy image of the bonded region, in which the dark areas indicate bonding. In the following, the top and bottom fiber are labeled as, respectively, + and -.}
	\label{fig:init_char}
\end{figure}

The full-field single fiber hygro-expansion methodology is naturally extended to (isolated) inter-fiber bonds. Orthogonal inter-fiber bonds are cut from the prepared sparse networks and clamped onto a specifically designed sample holder using two 50 $\mu$m nylon wires to minimize rigid body translations and to keep the inter-fiber bond inside the microscopic FOV, whereby the wires still allow for fiber swelling and (reduced) rotation and bending, see Figure \ref{fig:init_char} (a). As the hygro-mechanics of inter-fiber bonds is more difficult to interpret than that of single fibers, it is necessary to know the initial 3D geometry of each tested inter-fiber bond, i.e., the thickness and width of the fibers along their length, as well as the wrap around angle, contact surface and fiber-to-fiber bond angle. Furthermore, the initial 3D geometry is also required for the bi-layer laminate model describing the mechanics of the bonded area, see Subsection \ref{subs:lam_mod} below. To enable the measurement of the 3D geometry of the inter-fiber bonds, the specifically designed sample holder is schematically shown in Figure \ref{fig:init_char} (a), consisting of two thin (\textsuperscript{$\sim$}100 µm) glass slides, spaced \textsuperscript{$\sim$}300 µm from each other, glued to a rigid PMMA frame.
Placing a mirror (e.g. silicon wafer) underneath the prepared specimen allows to visualize two aligned topographies of the front and back of the inter-fiber bond as proposed by \cite{Maraghechi2023mirror} and shown in Figure \ref{fig:init_char} (b). The quality of the back topography is significantly less than the front due to the loss of brightness caused by the light rays being blocked by the inter-fiber bond, when the image is made via the mirror. The offset between the two topographies is known from the piezo-actuator inside the optical profilometer with reasonable accuracy (\textsuperscript{$\sim$}1 $\mu$m, determined by measuring wires of known diameter), allowing to extract a fiber thickness map as shown in Figure \ref{fig:init_char} (c). Computing the fiber width average height profile, shown in Figure \ref{fig:init_char} (d), allows for determining the degree of wrap around, given as $\phi_{-,1}$ and $\phi_{-,2}$ for, respectively, both sides of the bottom fiber, and $\phi_{+,1}$ and $\phi_{+,2}$ for the top fiber. The top and bottom fiber are labeled with, respectively, $+$ or $-$ in the following. Furthermore, the contact surface area is estimated using the methodology proposed by \cite{kappel2009novel}, where the dark areas in the polarization optical microscopy images, shown in Figure \ref{fig:init_char} (d), indicate the bonded regions. Finally, the fiber-to-fiber bond angle, given by $\alpha$ in Figure \ref{fig:init_char} (b) is also determined. In this work a total of four inter-fiber bonds are characterized and tested, and Table \ref{tab:character} shows the obtained characteristics for every inter-fiber bond, i.e. fiber thickness ($t$), and width ($w$), wrap around angle ($\phi$), estimated contact surface ($A$) with bonded fraction, and fiber-to-fiber bond angle ($\alpha$). Please note that, two inter-fiber bonds reveal a bond angle of 90$^o$, whereas two other bonds have a smaller angle, respectively, 71 and 78$^o$.
\begin{table}[H]
\centering
\caption{Inter-fiber bond characteristics: fiber thickness ($t$), fiber width ($w$), interface angle ($\phi_{1,2}$) of the top (+) and bottom (-) fiber, contact surface ($A$), and bond angle ($\alpha$)}
\label{tab:character}
	\resizebox{0.7\linewidth}{!}{\begin{tabular}{@{}ccccccccc@{}}
		\toprule
		\begin{tabular}[c]{@{}c@{}}inter-fiber\\ bond \end{tabular}  & \begin{tabular}[c]{@{}c@{}}$t_+$\\ {[}$\mu$m{]} \end{tabular} &  \begin{tabular}[c]{@{}c@{}}$t_-$\\ {[}$\mu$m{]} \end{tabular}  & \begin{tabular}[c]{@{}c@{}}$w_+$\\ {[}$\mu$m {]}\end{tabular}  & \begin{tabular}[c]{@{}c@{}}$w_-$\\ {[}$\mu$m{]} \end{tabular}  & \begin{tabular}[c]{@{}c@{}}$\phi_{1,+}$,  $\phi_{2,+}$\\ {[}\textsuperscript{o}{]} \end{tabular}  & \begin{tabular}[c]{@{}c@{}}$\phi_{1,-}$,  $\phi_{2,-}$\\ {[}\textsuperscript{o}{]} \end{tabular}  & \begin{tabular}[c]{@{}c@{}}$A$\\ {[}$\mu$m$^2${]} \end{tabular}  & \begin{tabular}[c]{@{}c@{}}$\alpha$\\ {[}\textsuperscript{o}{]} \end{tabular}  \\ \midrule
		1                & 2.8 & 3.2 & 57 & 53 & 2.6, 6.4 & 4.2, 6.5 & 2533 (84\%) & 90\\
		2                & 3.2 & 4.1 & 25 & 28 & 11.0, 8.7 & 11.2, 14.4 & 623 (89\%) & 90\\
		3                & 1.5 & 4.2 & 44 & 62 & 1.5, 2.4 & 2.6, 5.8 & 1140 (76\%) & 79\\
		4                & 1.5 & 5.1 & 61 & 64 & 1.7, 0.9 & 3.1, 3.8 & 3475 (89\%) & 78\\ \bottomrule
	\end{tabular}}
\end{table}

\subsection{Acquisition and GDHC parameter determination}
\begin{figure}[t!]
	\centering
	\includegraphics[width=0.5\textwidth]{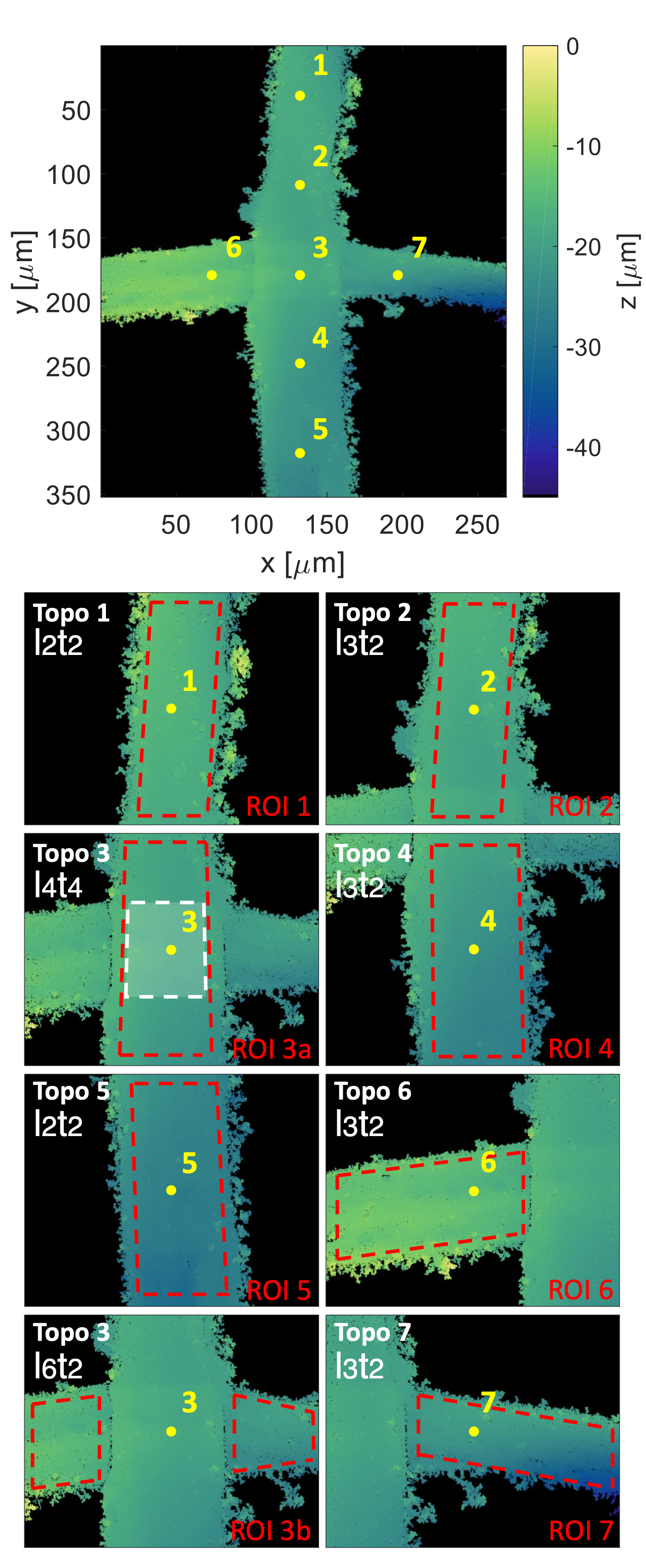}
	\caption{Stitched image of an inter-fiber bond with the location of the center of seven separate topographies depicted below. The seven ROIs used in the GDHC are highlighted in the separate images and the optimal order of the polynomial field is added, where, e.g., "l3t2" indicates that a third and second order polynomial used in, respectively, the longitudinal and transverse direction for the correlation. Please note that image 3 is correlated twice with different ROIs, since it contains both the bottom and top fiber. The central area of ROI 3a is highlighted in white which indicates the bonded area used in Figure \ref{fig:strains_bend} below.}
	\label{fig:ROI_bond}
\end{figure}

After the initial characterization, the specimen is patterned and tested following the single fiber procedure described in \citep{vonk2020robust}. Each inter-fiber bond is subjected to a RH cycle of 30$-$90$-$30\% for four cycles, with a constant rate of 0.1\%/min and a plateau of 5 hours at 90 or 30\%, resulting in a total time per inter-fiber bond of 120 hours. The small rate is necessary to be able to capture seven parallel 55$\times$ topographies (with a FOV of 147$\times$111 $\mu$m\textsuperscript{2}) of the inter-fiber bond with its four connected "arms" with sufficient detail for the GDHC. Please note that the inter-fiber bond 2 is imaged using a magnification of 100$\times$ (FOV of 81$\times$61 $\mu$m\textsuperscript{2}) because the width of both fibers is significantly smaller than the other inter-fiber bonds, see Table \ref{tab:character}. As an example, the location of the seven topographies (topo) are shown in Figure \ref{fig:ROI_bond}. An overlap of 25\% of, respectively, the height and width of the topography is chosen.
Two different correlation strategies can be used to obtain the full-field behavior of the tested inter-fiber bonds: correlating the stitched images and extract continuous deformation fields, or correlating the seven images separately, resulting in dis-continuous deformation fields. Prior analysis revealed that the stitched images cannot trivially be used in the GDHC algorithm to obtain a continuous strain field, as the stitching entails sub-pixel artifacts into the images which would result in artificial strain spots. Nevertheless, a high strain precision along the full inter-fiber bond surface is required since longitudinal strains of around 0.2\% are expected \citep{vonk2021full}. To achieve sufficiently precise strains, each of the topographies will be correlated individually in each of the eight regions-of-interest (ROIs), after which the resulting individual strain fields are stitched into continuous strain profiles. To this end, ROIs 1$-$3a$-$5 are combined for the top fiber strain profiles and ROIs 6, 3b and 7 for the bottom fiber strain profile. \\ \indent
In \citep{vonk2020robust}, it was found that a second order polynomial in both fiber length and width direction offered the best kinematic regularization for describing single fiber swelling. However, as significantly larger strain gradients are expected in and around the bonded area, a higher order polynomial description is required to properly describe the displacement fields at and near the bonded area. Therefore, a polynomial order analysis, as done by \cite{neggers2014direct} and \cite{hoefnagels2022accurate}, is conducted. The strain gradients in the fiber's transverse direction are expected to be similar to single fibers, whereas major variations are expected along the fiber's longitudinal direction. Hence, the eight ROIs are correlated using a 2\textsuperscript{nd}$-$5\textsuperscript{th} order polynomial description in the transverse and 2\textsuperscript{nd}$-$7\textsuperscript{th} order in the longitudinal direction. The correlation residual and strain fields for every polynomial combination have been assessed and it was found that for ROI 1 and 5 a second order polynomial field in both direction suffices. For ROI 2, 4, 6, and 7 (closer to the bonded area) a third order polynomial and second order in, respectively, longitudinal and transverse direction appears to be optimal. ROI 3a, containing the top fiber, requires a fourth order polynomial in both directions, while the bottom fiber in ROI 3b, is best correlated using a sixth and second order polynomial in, respectively, longitudinal and transverse direction. The polynomial orders used for the correlation of each ROI are added to Figure \ref{fig:ROI_bond}, where, for instance, the code "l3t2" indicates a third and second order polynomial in, respectively, longitudinal and transverse direction. ROI 3b consists of two separate ROIs that are correlated at once with a continuous "l6t2" displacement field, thereby taking into account the relative displacement between the fiber segments left and right of the bonded area. The resulting displacement field provides, by means of interpolation, an estimate of the longitudinal (horizontal) strain field inside the bonded area at the fiber-to-fiber interface surface, as detailed below.

\subsection{Data processing}
\begin{figure}[H]
	\centering
	\includegraphics[width=1\textwidth]{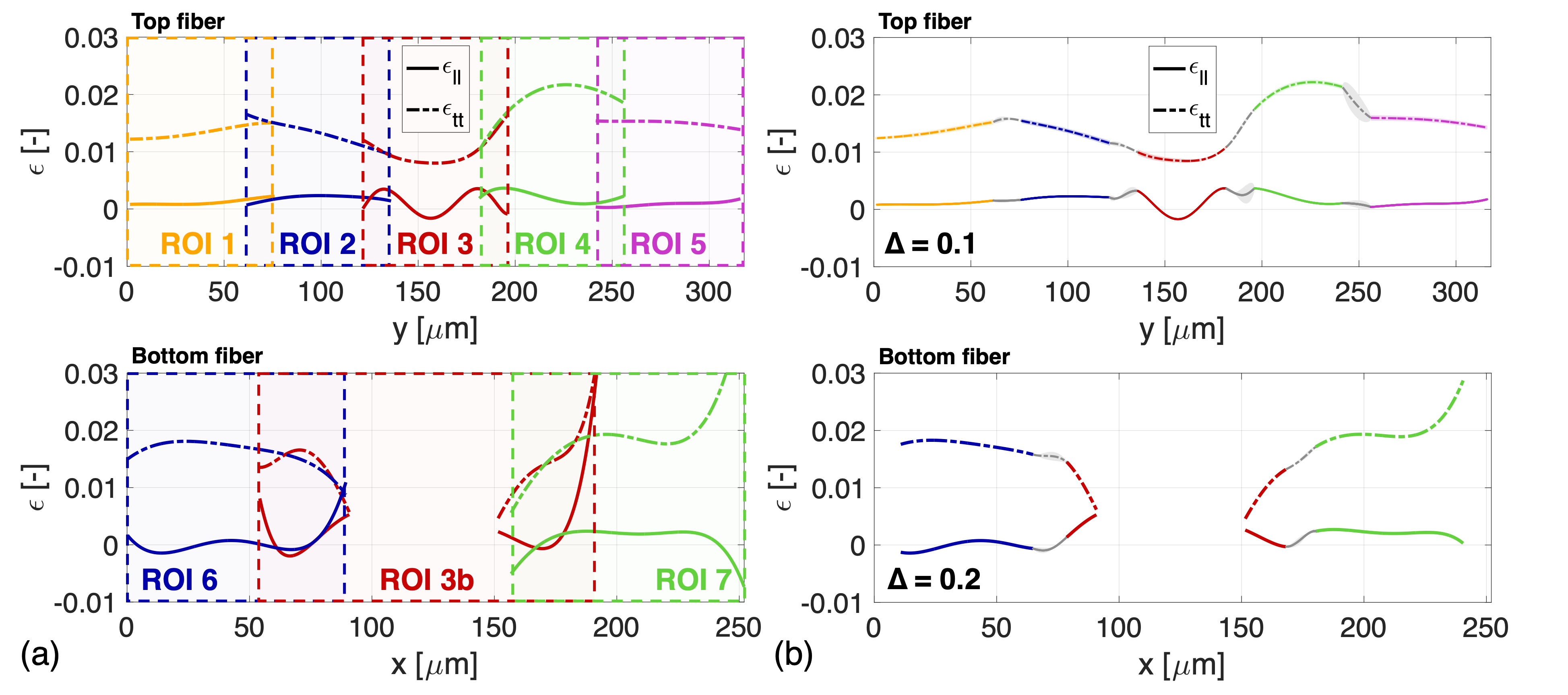}
	\caption{The fiber width averaged transverse and longitudinal strain increase from 30 to 90\% RH of the top and bottom fiber for each correlated ROI, with (a) the unprocessed strain trend of each ROI correlation, showing clear mismatches in the overlapping regions, and (b) continuous strain trends, with reduced errors in the overlapping regions, by means of linear scaling, for two different values of the width $\Delta$ of a band of data points spanning the edge of the overlapping region. A factor $\Delta$ of 0.1 represents a band of 10\% of the overlapping region which is neglected in the average trend determination in the overlapping regions.}
	\label{fig:strains_ROI}
\end{figure}

The longitudinal and transverse hygro-expansion, averaged over the fiber width, for an RH increase of 30 to 90\% (cycle 2) of the top and bottom fiber for each ROI is given in Figure \ref{fig:strains_ROI} (a). The global curves are obvious, but clear mismatches emerge in the overlapping regions. This is driven by the fact that polynomials (especially higher order) tend to fluctuate near the edges of the ROI \citep{neggers2014direct}.
To remedy this issue, in the overlap area, the strain values are stitched by linearly scaling the strains of one of the two adjacent ROIs from 1 to 0, while in the adjacent ROI a reverse scaling from 0 to 1 is used, and adding both scaled strains, in the spirit of a partition of unity to yield the stitched strains. Additionally, an error for every average strain point is determined to estimate the stitching error. However, the outer edges of e.g. ROI 3b in Figure \ref{fig:strains_ROI} (a) reveal such large non-physical deviations, i.e. these faulty regions are not adequately reduced by the linear scaling and therefore need to be discarded. Hence, a factor $\Delta$ is introduced, representing a band around the border of each overlapping region which will not be taken into account in combining the strain curves, e.g., a factor of $\Delta$ = 0.1 would neglect a band of 10\% of the width of the overlapping area. By neglecting these boundary regions the expected accuracy of the measurement is further improved. After comprehensive analysis, a $\Delta$ of 0.1 and 0.2 provides the best results for, respectively the top and bottom fiber for all tested inter-fiber bonds, and the resulting strains are plotted in Figure \ref{fig:strains_ROI} (b). The larger factor $\Delta$ for the bottom fiber is attributed to ROI 3b being correlated with a higher order polynomial field compared to the top fiber, hence the errors around the edges of ROI 3b span a wider region. In the resulting stitched strain profiles, no discontinuities can be spotted (see, e.g., Figure \ref{fig:strains_ROI} (b)) and a smooth curve is obtained for all strain profiles of all inter-fiber bonds. Please note that, while the overlapping region of the topographies for every inter-fiber bond and $\Delta$ are equal for every inter-fiber bond, the size of the overlapping regions in the strain curves differ somewhat because the bounds of the correlated ROIs are not the same. 

\section{Results and Discussion}

\subsection{Full-field hygro-expansivity}
\label{ssec:hygro}
\begin{figure}[H]
	\centering
	\includegraphics[width=\textwidth]{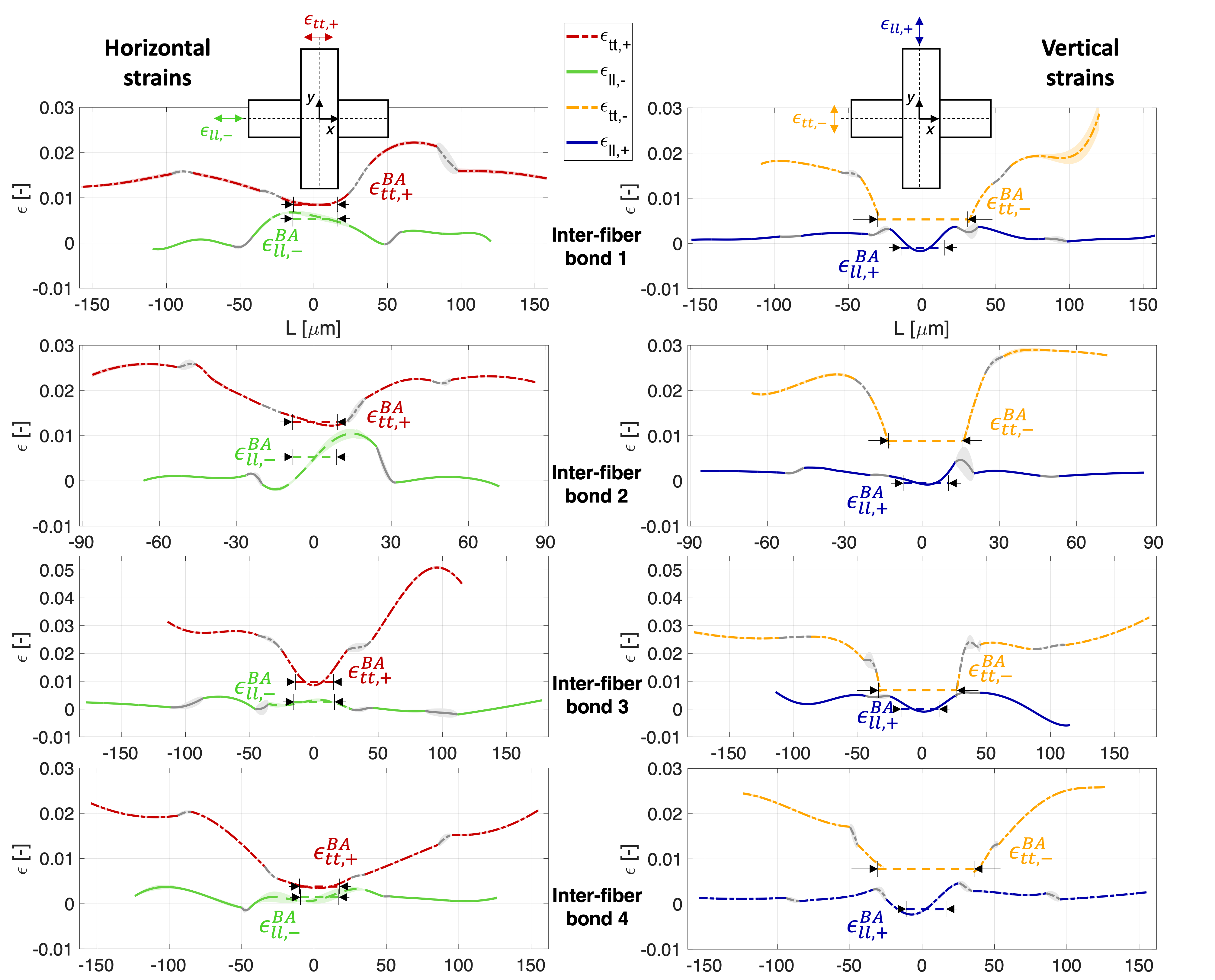}
	\caption{Fiber-width averaged longitudinal and transverse strain of, respectively, the top and bottom fiber along the fiber lengths for an RH increase of 30 to 90\% of the inter-fiber bonds. The curves shown are the average curve of cycles 2$-$4 with its standard deviation shown by the shaded band which is often too thin to see. The transverse strain of the top fiber ($\epsilon_{tt,+}$) and the longitudinal strain of the bottom fiber ($\epsilon_{ll,-}$) are in the horizontal direction but at different heights of the bond, and the longitudinal strain of the top fiber ($\epsilon_{ll,+}$) and transverse strain of the bottom fiber ($\epsilon_{tt,-}$) are in the vertical direction. Similar responses are found for each inter-fiber bond, indicating qualitatively reproducible hygro-mechanical behavior. For each inter-fiber bond, (i) the transverse strain of the top and bottom fiber decreases in and close to the bonded area, (ii) a decrease of the longitudinal strain of the top fiber ($\epsilon_{ll,+}$) is found and (iii) an increase of the longitudinal strain of the bottom fiber ($\epsilon_{ll,-}$) due to the transverse strain transferal of the top fiber. The average bond strain ($\epsilon^{BA}$) is depicted in each graph by the horizontal dashed lines.}
	\label{fig:strains_all}
\end{figure}

To explore the full-field nature of the resulting data, as given for inter-fiber bond 1 in Figure \ref{fig:strains_ROI} (b), the fiber-width averaged longitudinal and transverse strain of, respectively, the top and bottom face along the fiber lengths for an RH increase from 30 to 90\% of all inter-fiber bonds are determined. The horizontal and vertical strain curves of all inter-fiber bonds are presented in Figure \ref{fig:strains_all}, where the transverse strain of the top fiber ($\epsilon_{tt,+}$) and the longitudinal strain of the bottom fiber ($\epsilon_{ll,-}$) are oriented in the horizontal direction, whereas the longitudinal strain of the top fiber ($\epsilon_{ll,+}$) and the transverse strain of the bottom fiber ($\epsilon_{tt,-}$) are both oriented in the vertical direction, both at different heights through the thickness of the inter-fiber bond. Note that the centers of the bonded area of the top and bottom fiber are centrally aligned. The average strain profiles, based on wetting cycles 2$-$4 (because the first wetting cycle shows a significant release of dried-in strain as shown in Figure \ref{fig:strains_bend} (a) below), is shown together with its standard deviation, that is visualized by the (often very thin) band around the curves. \\ \indent
First, the hygroscopic behavior of each RH cycle is reproducible, because the resulting scatter for each strain component is low, highlighting the accuracy of the used GDHC method. Furthermore, when considering the strains in the bonded region, a repeating pattern is visible for every inter-fiber bond: (i) the transverse strain of both fibers ($\epsilon_{tt,+}$ and $\epsilon_{tt,-}$) decreases when approaching the bonded area, which is attributed to the low longitudinal strain of the bonded fiber, combined with a larger longitudinal stiffness \citep{magnusson2013numerical, czibula2021transverse}, constraining the transverse hygro-expansion, (ii) the longitudinal strain of the top fiber shows compression in the bonded area, which is due to the out-of-plane bending of the bond, as studied numerically by \cite{brandberg2020role} and analyzed in more detail below, and (iii) the longitudinal strain of the bottom fiber ($\epsilon_{ll,-}$), with the interpolated strain field depicted by the dashed line, increases when approaching the bonded area from both sides. This suggests that the large transverse strain of the top fiber stretches the bottom fiber in its longitudinal direction, as experimentally observed by \cite{nanko1995mechanisms}. \\ \indent
These observations are strong evidence that the hygro-mechanical behavior of the isolated inter-fiber bonds is qualitatively consistent (and the measurement method reproducible) even for inter-fiber bonds consisting of deviating fiber geometries and fiber-to-fiber bond angles (as seen for inter-fiber bonds 3 and 4). Note that for inter-fiber bond 3, five and three topographies have been used to image, respectively, the bottom and top fiber instead of vice versa (as done for the other inter-fiber bonds). This was necessary to accommodate the severe fiber surface roughness at the far ends of the vertical fiber.\\ \indent
The average strain in the bonded area (i.e. the average bond strain) is determined and depicted in each graph in Figure \ref{fig:strains_all} by $\epsilon_{ll,+}^{BA}$, $\epsilon_{tt,+}^{BA}$, $\epsilon_{ll,-}^{BA}$, and $\epsilon_{tt,-}^{BA}$. The magnitude of $\epsilon_{ll,+}^{BA}$, $\epsilon_{tt,+}^{BA}$, and $\epsilon_{ll,-}^{BA}$ is determined by considering the central area of the strain field in the bond, i.e. a band of 25\% around the strain field is excluded, as depicted in Figure \ref{fig:strains_all}. The transverse average bond strain of the bottom fiber ($\epsilon_{tt,-}^{BA}$) is determined as follows. Considering that the longitudinal strain of the top fiber constrains and reduces $\epsilon_{tt,-}$ at the bonded area, $\epsilon_{tt,-}^{BA}$ should be equal to or lower than the value of $\epsilon_{tt,-}$ just next to the bond. Hence, $\epsilon_{tt,-}^{BA}$ is taken as the average of the closest measurement value of $\epsilon_{tt,-}$ left and right of the bonded area. Note that each of the bond strain values, $\epsilon^{BA}$, is taken as the average of cycles 2$-$4. In summary, the obtained full-field hygro-expansivity data allows for accurate longitudinal and transverse strains in and around the bonded region. Moreover, the full-field data enables identification of the average bond strains at the top fiber surface and the fiber-to-fiber interface in longitudinal and transverse direction. In order to better understand the above-described strain curves, the inter-fiber bond hygro-mechanics need proper investigation as elaborated next.

\subsection{Inter-fiber bond hygro-mechanics}
\begin{figure}[H]
	\centering
	\includegraphics[width=0.5\textwidth]{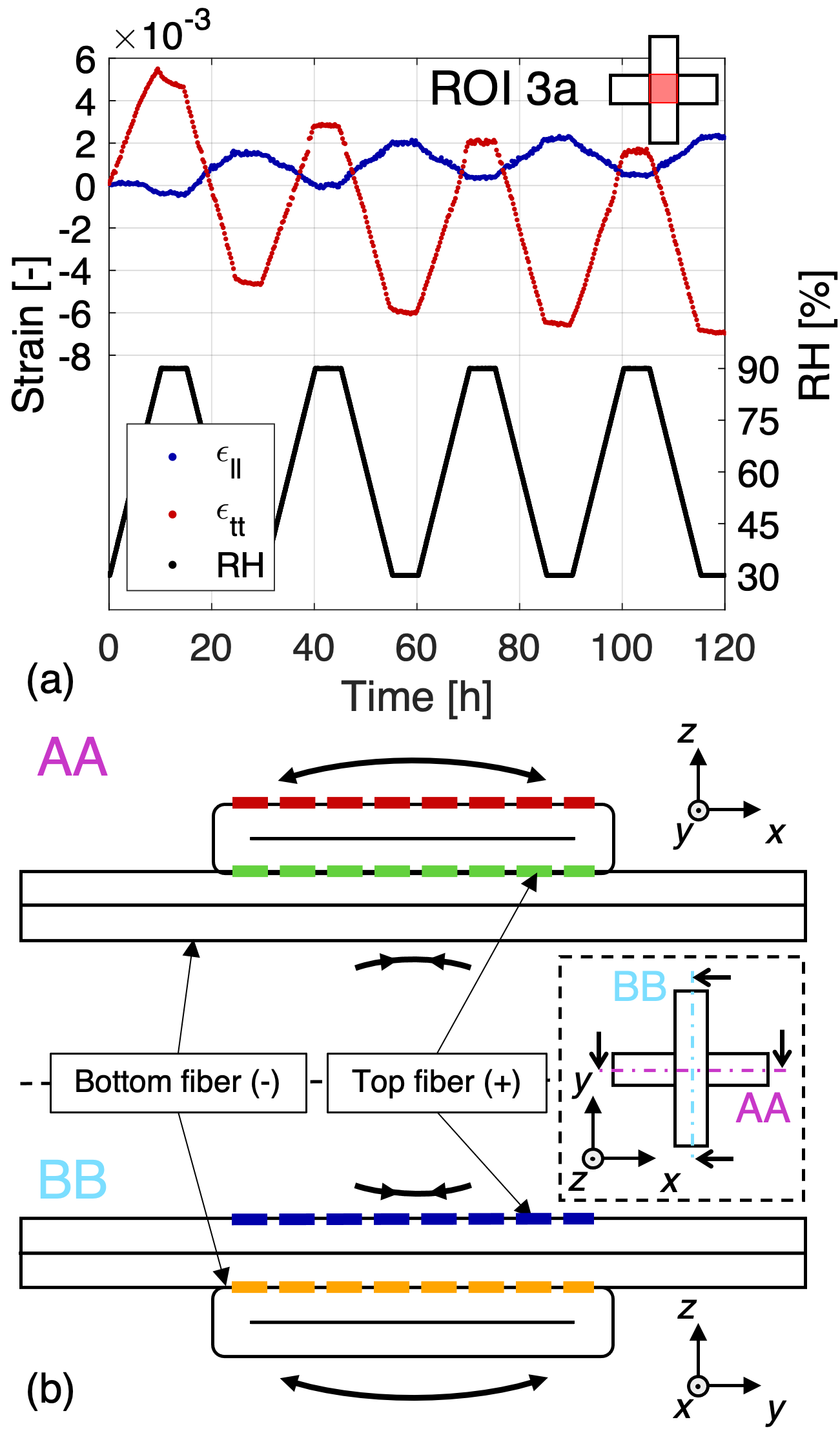}
	\caption{Inter-fiber bond hygro-mechanics, with (a) the field-averaged longitudinal and transverse hygro-expansion in the bonded area of ROI 3a (highlighted white in Figure \ref{fig:ROI_bond}), and the measurement of the RH, which is controlled over time, for inter-fiber bond 1. During the increase of RH, the increase of the transverse strain is accompanied by a decrease of the longitudinal strain, a phenomenon that is driven by the significantly larger transverse (compared to longitudinal) hygro-expansivity of the paper fibers, resulting in out-of-plane bending of the inter-fiber bond upon wetting or drying. In (b), cross-section AA shows downward bending because the top fiber expands more in its transverse direction than the bottom fiber in its longitudinal direction. Vice versa for cross-section BB, resulting in upward bending. The colors in the cross-sections correspond to the strain in longitudinal and transverse direction at, respectively, the top and interface surface, which are in the same direction for every cross-section.}
	\label{fig:strains_bend}
\end{figure}

Figure \ref{fig:strains_bend} (a) shows the evolution of the field-averaged longitudinal and transverse hygro-expansion in the bonded region of ROI 3a (top fiber) of inter-fiber bond (highlighted in white in Figure \ref{fig:ROI_bond}), together with the RH that is controlled over time. 
\noindent
The overall transverse hygro-expansion response is similar to the single fiber responses displayed in Figure \ref{fig:strains_fiber}, i.e. an overall release of irreversible (dried-in) strain. However, the transverse hygro-expansivity is significantly lower compared to the single fibers found in Figure \ref{fig:strains_fiber}, implying a reduction of the transverse strain of the top fiber by the lower longitudinal strain of the bottom fiber, which are (globally) oriented in the same direction for inter-fiber bond 1 and 2 (with a bond angle of 90\textsuperscript{o}). Moreover, for the wetting cycles 2$-$4, the longitudinal strain decreases for increasing RH, and vice versa, which contradicts the general explanation based on swelling or shrinkage during, respectively, moisture uptake or release. The only logical explanation for this negative longitudinal hygro-expansion at the top fiber surface in the bonded area during moisture uptake is upward (out-of-plane) bending, which is driven by the significant difference in longitudinal and transverse hygro-expansivity and stiffness of the fibers constituting the bond \citep{vonk2021full, magnusson2013numerical}, as also visualized by the 3D bond model proposed by \cite{brandberg2020role}. Please note that all inter-fiber bonds reveal the same curve as given in Figure \ref{fig:strains_bend} (a) and hence display the described bending behavior, as visualized by the decrease in longitudinal strain for all inter-fiber bonds in Figure \ref{fig:strains_all}) \\ \indent
Two schematic cross-sections of the inter-fiber bond are shown in Figure \ref{fig:strains_bend} (b), where AA and BB refer to, respectively, the horizontal and vertical cross-section. Cross-section BB is considered for understanding the behavior found in Figure \ref{fig:strains_bend} (a). During moisture uptake, the bottom fiber swells significantly more in transverse direction than the top fiber in longitudinal direction, which causes upward bending. Reversibly, in cross-section AA, the large transverse strain of the top fiber and low longitudinal strain of the bottom fiber results in downward bending. Note that the colors in the cross-sections are linked to the strains through the thickness of the bond, i.e. blue and red corresponds to, respectively the longitudinal and transverse strain at the top surface of the top fiber and green and orange to, respectively, the longitudinal and transverse strain of the bottom fiber at the interface surface, which are both in the same direction for the separate cross-sections. \\ \indent
One can question if this out-of-plane bending also occurs for inter-fiber bonds located inside a paper sheet, where they are severely constrained by other fibers. Therefore, an analytical bending model for a thin laminate based on classical laminate theory is proposed to rationalize this bending behavior. The model can subsequently be compared to the experimentally obtained average bond strains. If this model adequately captures the bond mechanics, it would allow to extrapolate the strain profile through the complete thickness of the inter-fiber bond, even at the bottom which was not characterized. Additionally, the model (with different boundary conditions) may also be used for the inter-fiber bond hygro-mechanics inside a paper sheet, where minimal (or zero) bending is possible, thereby analyzing the transverse strain transferal at the sheet level. This bi-layer laminate model is explored next. 

\subsection{Bi-layer laminate model describing the bonded area}
\label{subs:lam_mod}
\begin{figure}[H]
	\centering
	\includegraphics[width=0.5\textwidth]{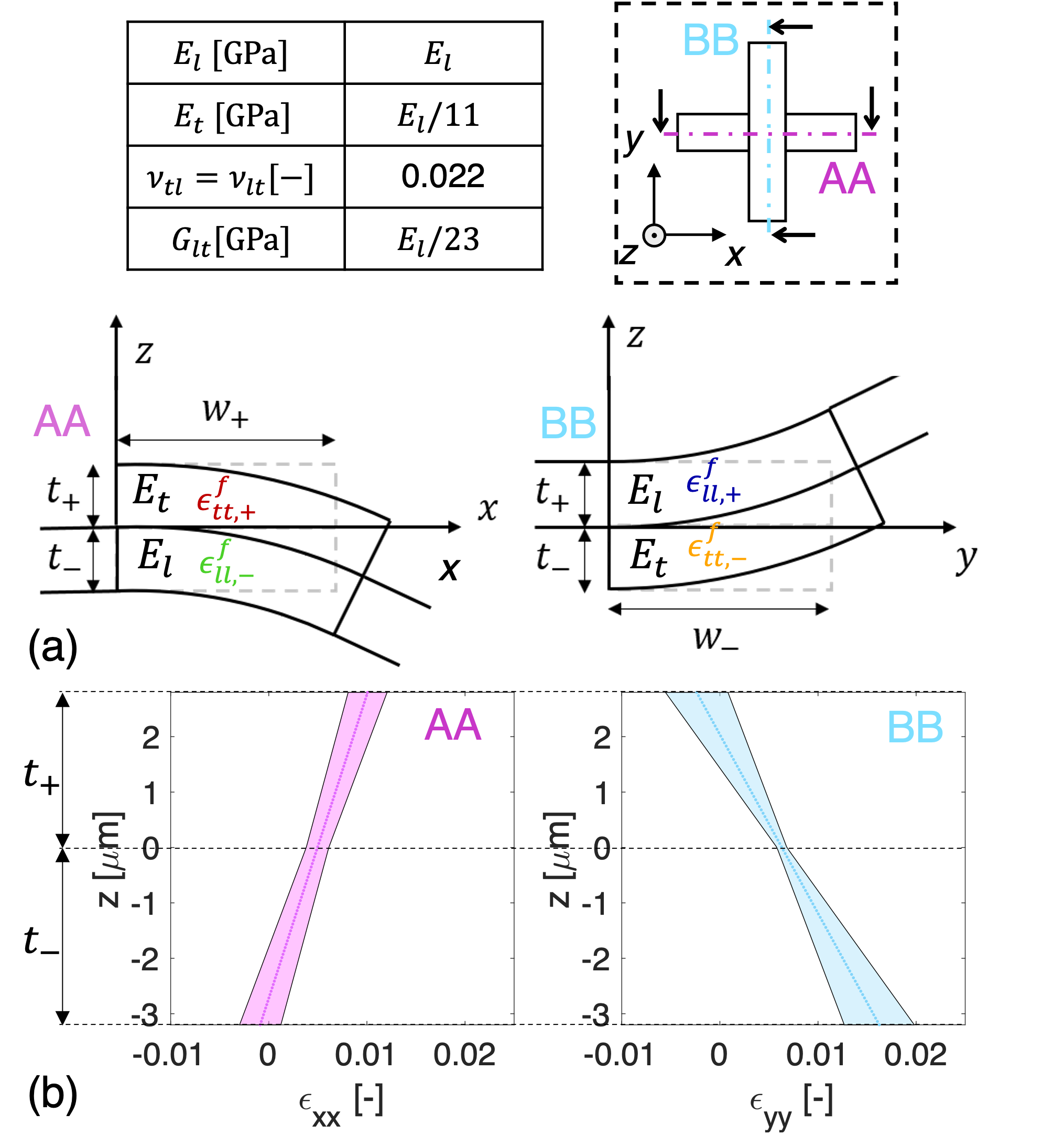}
	\caption{Bi-layer laminate model used to describe the bending deformation in the bonded area, with (a) the horizontal and vertical cross-section AA and BB of the inter-fiber bond with described bending deformation and the used material parameters, which are obtained from \citep{magnusson2013numerical}. The fiber dimensions and hygro-expansion of the freestanding fiber segments ($\epsilon^f$) used in the model are given in, respectively, Table \ref{tab:character} and \ref{tab:hygro}. (b) The predicted linear strain curves though the thickness of inter-fiber bond 1 with the confidence area marking the uncertainty propagation of uncertainties of the input of the model, such as the hygro-expansion and dimensions given in Table \ref{tab:character} and\ref{tab:hygro}.}
	\label{fig:lam_mod}
\end{figure}

Because the width of the fiber is significantly larger than the thickness, the classical laminate theory for thin plates can be adopted to describe the bending deformation in the bonded area \citep{pister1959elastic, reissner1961bending}. The orientation of the two plates are rotated in the (longitudinal-transverse) fiber plane to each other with the fiber-to-fiber bond angle, $\alpha$, as given in Table \ref{tab:character}. $\alpha$ is 90\textsuperscript{o} for inter-fiber bond 1 and 2, therefore, a perpendicular fiber bond geometry will be assumed in the model. Moreover, the strain over the thickness of the laminate, which usually is modeled with an elastic and a thermal component in standard laminate theory, is substituted by an elastic and hygroscopic strain ($\epsilon^H$),
\begin{equation}
	\epsilon(z) = \epsilon^E + \epsilon^H,
	\label{eq:strains}
\end{equation}
Assuming a linear dependency of the strain though the thickness to the curvature ($\kappa$) of the laminate results in,
\begin{equation}
	\epsilon(z) = \epsilon_{0}-\kappa z,
	\label{eq:prof}
\end{equation}
in which $\epsilon_0$ represents the strain value at the interface between the two fibers (defined as $z = 0$). Combining Equations \ref{eq:strains} and \ref{eq:prof}, the following equation for the elastic strain is found:
\begin{equation}
	\epsilon^E(z) = \epsilon_0 - \kappa z - \epsilon^H, 
	\label{eq:temp}
\end{equation}
where $\epsilon^H$ can be obtained from the measurements of the freestanding segments of the fibers at the outer ends, denoted as $\epsilon^f$. The strain at the interface ($\epsilon_0$) and curvature of the laminate ($\kappa$) are yet to be determined in order to obtain the strain profile through the thickness of the bonded area. \\ \indent
Figure \ref{fig:lam_mod} (a) displays two schematic cross-sectional representations of the bending problem in horizontal (AA) and vertical direction (BB). In cross-section BB, the top fiber's longitudinal and the bottom fiber's transverse direction are in $y$-direction (resulting in upward bending because the transverse hygro-expansivity is always larger than the longitudinal), and vice versa in $x$-direction for cross-section AA (resulting in downward bending). Note that perfect bonding between the two fibers is assumed in this model.
\noindent
The goal is to determine $\epsilon_0$ and $\kappa$ that accommodate the linear strain profiles through the thickness of the bonded region (Equation \ref{eq:temp}). This requires solving the force and moment balance equations. Assuming free expansion ($F = 0$ and $M = 0$), the following equations are obtained in $x$-direction,
\begin{equation}
	\label{eq:FM}
	\begin{split}
		F_{xx} = w_+\int_{t_-}^{0}\sigma_{xx,-}dz+w_+\int_{0}^{t_+}\sigma_{xx,+}dz = 0, \\ M_{xx} = w_+\int_{t_-}^{0}z\sigma_{xx,-}dz+w_+\int_{0}^{t_+}z\sigma_{xx,+}dz = 0,
	\end{split}
\end{equation}
in which $w$ is the width, $t_-$ and $t_+$ the thickness of the top and bottom layer forming the laminate, and $\sigma_{xx,-}$ and $\sigma_{xx,+}$, the horizontal stress in the top and bottom layer, respectively, which are related to the strain through the constitutive behavior for orthogonal inter-fiber bonds:

\begin{equation}
	\label{eq:stress}
	\begin{aligned}
		\sigma_{xx,-} ={ } & C_{11}(\epsilon_{xx} - \epsilon_{xx,-}^H) + C_{12}(\epsilon_{yy} - \epsilon_{yy,-}^H) + C_{16}(\epsilon_{xy} - \epsilon_{xy,-}^H),\\
		\sigma_{xx,+} =	{ }& C_{12}(\epsilon_{yy} - \epsilon_{yy,+}^H)+C_{22}(\epsilon_{xx} - \epsilon_{xx,+}^H) + C_{26}(\epsilon_{xy,+} - \epsilon_{xy}^H),
	\end{aligned}
\end{equation}
with $\epsilon_{xx} = \epsilon_{x,0} - \kappa_{xx} z$, $\epsilon_{yy} = \epsilon_{y,0} - \kappa_{yy} z$, $\epsilon_{xy} = \epsilon_{xy,0} - \kappa_{xy} z$, and $\epsilon_{xx,-}^H=\epsilon_{ll,-}^f$, $\epsilon_{yy,-}^H=\epsilon_{tt,-}^f$, $\epsilon_{xy,-}^H=\epsilon_{lt,-}^f$, $\epsilon_{yy,+}^H=\epsilon_{ll,+}^f$, $\epsilon_{xx,+}^H=\epsilon_{tt,+}^f$, $\epsilon_{xy}^H=\epsilon_{lt,+}^f$ and the stiffness coefficients, $C$, are
\begin{equation}
	C_{11} = \frac{E_l}{1-\nu_{lt}\nu_{tl}}, \quad C_{12} = \frac{\nu_{tl}E_l}{1-\nu_{lt}\nu_{tl}}, \quad C_{22} = \frac{E_t}{1-\nu_{lt}\nu_{tl}}, \quad C_{16} = C_{26} = 0, \quad C_{66} = G_{lt},
	\label{eq:stiff}
\end{equation}
in which $E$ is the Young's modulus and $\nu$ the Poisson's ratio, which are given in the table in Figure \ref{fig:lam_mod}. \\ \indent 
Table \ref{tab:character} provides the geometries of the top and bottom fiber, while Table \ref{tab:hygro} gives the average longitudinal ($\epsilon_{ll}^f$), transverse ($\epsilon_{tt}^f$) and shear strain increase ($\epsilon_{lt}^f$) with standard deviation for an RH increase of 30 to 90\% averaged over cycles 2$-$4. The remaining balance equations for $F_{yy}$, $F_{xy}$, $M_{yy}$ and $M_{xy}$, in which $\sigma_{xx}$ in Equation \ref{eq:FM} is replaced by, $\sigma_{yy}$ and $\sigma_{xy}$ for the $yy$ and $xy$ component of the force and moment, respectively \citep{reissner1961bending}, resulting in six equations that are used to determine the six unknown constants $\epsilon_{0,xx}$, $\epsilon_{0,yy}$, $\epsilon_{0,xy}$, $\kappa_{xx}$, $\kappa_{yy}$, and $\kappa_{xy}$. Substitution of these constants into Equation \ref{eq:prof} results in three equations describing a linear strain profile though the thickness of the bond in each direction, i.e. $\epsilon_{xx} = \epsilon_{0,xx}+\kappa_{xx}z$, $\epsilon_{yy} = \epsilon_{0,yy}+\kappa_{yy}z$, and $\epsilon_{xy} = \epsilon_{0,xy}+\kappa_{xy}z$, of which the horizontal $\epsilon_{xx}(z)$ and vertical strain curve $\epsilon_{yy}(z)$ as a function of the thickness of inter-fiber bond 1 are displayed in Figure \ref{fig:lam_mod} (b). The uncertainty in curvature ($\kappa$) and interface strain ($\epsilon_0$) are determined by means of propagating the uncertainties of the input parameters of the model, which are the uncertainty in hygro-expansivity of the freestanding segments considering cycles 2$-$4 and the fiber dimensional measurement uncertainties. Resulting in four lines (i.e. $\epsilon_1 = (\epsilon_0+d\epsilon_{0})+(\kappa+d\kappa)z$, $\epsilon_2 = (\epsilon_0+d\epsilon_{0})+(\kappa-d\kappa)z$, $\epsilon_3 = (\epsilon_0-d\epsilon_{0})+(\kappa+d\kappa)z$, and $\epsilon_4 = (\epsilon_0-d\epsilon_{0})-(\kappa-d\kappa)z$), of which the outer limits of the enclosed area reflects the border of the confidence area as depicted in Figure \ref{fig:lam_mod} (b). \\ \indent
Note that the above-described equations are, for now, only applicable to orthogonal inter-fiber bonds (bonds 1 and 2). In order to make them applicable for non-orthogonal inter-fiber bonds, i.e. bonds 3 and 4, the hygro-expansivities and stiffness coefficients of one layer require rotation relative to the other layer \citep{reissner1961bending}. This is done by rotating the bottom layer, for which the rotated hygro-expansion ($\hat{\bm{\epsilon}}^H_-$) is determined by:
\begin{equation}
	\begin{bmatrix} \hat{\epsilon}_{xx,-}^H \\ \hat{\epsilon}_{yy,-}^H \\ \hat{\epsilon}_{xy,-}^H \end{bmatrix} = \begin{bmatrix} c^2 & s^2 & cs\\ s^2 & c^2 & -cs \\ -2cs & 2cs & c^2-s^2 \end{bmatrix} \begin{bmatrix} \epsilon_{ll,-}^H \\ \epsilon_{tt,-}^H \\ \epsilon_{lt,-}^H \end{bmatrix},
	\label{eq:rotated_h}
\end{equation}
in which $c$ and $s$ represent $\cos(90-\alpha)$ and $\sin(90-\alpha)$ respectively, with $\alpha$ of course being the fiber bond angle given in Table \ref{tab:character}. The non-rotated hygro-expansions of the bottom layer ($\bm{\epsilon}^H_-$) in Equation \ref{eq:stress} are replace rotated hygro-expansions ($\hat{\bm{\epsilon}}^H_-$) determined using Equation \ref{eq:rotated_h}. The stiffness coefficients of the bottom layer are rotated relative to the top layer following the tensoral rotation equations given by \cite{reissner1961bending}. To this end, $C_{16}$ and $C_{26}$ (Equation \ref{eq:stiff}) become non-zero, in contrast to orthogonal inter-fiber bonds. Equation \ref{eq:FM}, which now includes the bond angle, can again be solved to find the linear strain relations through the thickness of the non-orthogonal bonds.
\begin{table}[t!]
	\centering
	\caption{The average longitudinal ($\epsilon_{ll}$), transverse ($\epsilon_{tt}$) and shear ($\epsilon_{lt}$) hygro-expansivity with standard deviation of the freestanding segments of the bottom (-) and top (+) fiber of the tested inter-fiber bonds determined from the wetting slope from 30 to 90\% RH and averaged over cycles 2$-$4.}
	\label{tab:hygro}
	\resizebox{0.5\linewidth}{!}{\begin{tabular}{@{}cccc@{}}
			\toprule
			\begin{tabular}[c]{@{}c@{}}inter-fiber\\ bond \end{tabular}  &  
			\begin{tabular}[c]{@{}c@{}}$\epsilon_{ll,-}^f$ \\ {[}-{]}\end{tabular}  & \begin{tabular}[c]{@{}c@{}}$\epsilon_{tt,-}^f$ \\ {[}-{]}\end{tabular} &
			\begin{tabular}[c]{@{}c@{}}$\epsilon_{lt,-}^f$ \\ {[}-{]} \end{tabular}  \\ \midrule
			1                & 0.0014$\pm$0.0006& 0.0186$\pm$0.0013 & -0.0004$\pm$0.0046 \\
			2                & 0.0011$\pm$0.0003 & 0.0245$\pm$0.0043 & -0.0016$\pm$0.0025 \\
			3                & 0.0017$\pm$0.0011 & 0.0265$\pm$0.0021 & -0.0095$\pm$0.0061 \\
			4                & 0.0021$\pm$0.0016 & 0.0239$\pm$0.0023 & -0.0006$\pm$0.0025 \\ \bottomrule \toprule
			\begin{tabular}[c]{@{}c@{}}inter-fiber\\ bond \end{tabular} & \begin{tabular}[c]{@{}c@{}}$\epsilon_{ll,+}^f$\\ {[}-{]} \end{tabular}  & \begin{tabular}[c]{@{}c@{}}$\epsilon_{tt,+}^f$\\ {[}-{]} \end{tabular} & \begin{tabular}[c]{@{}c@{}}$\epsilon_{lt,+}^f$\\ {[}-{]} \end{tabular} \\\midrule
			1				& 0.0011$\pm$0.0002 & 0.0142$\pm$0.0023 &  -0.0007$\pm$0.0007 \\
			2				& 0.0018$\pm$0.0002 & 0.0237$\pm$0.0027 &  -0.0050$\pm$0.0026 \\
			3				& 0.0010$\pm$0.0026 & 0.0370$\pm$0.0089 & -0.0041$\pm$0.0036 \\
			4				& 0.0013$\pm$0.0002 & 0.0181$\pm$0.0024 & -0.0007$\pm$0.0012 \\ \bottomrule 
	\end{tabular}}
\end{table}

\subsubsection{Comparison between bending model and experiments}
\begin{figure}[H]
	\centering
	\includegraphics[width=0.5\textwidth]{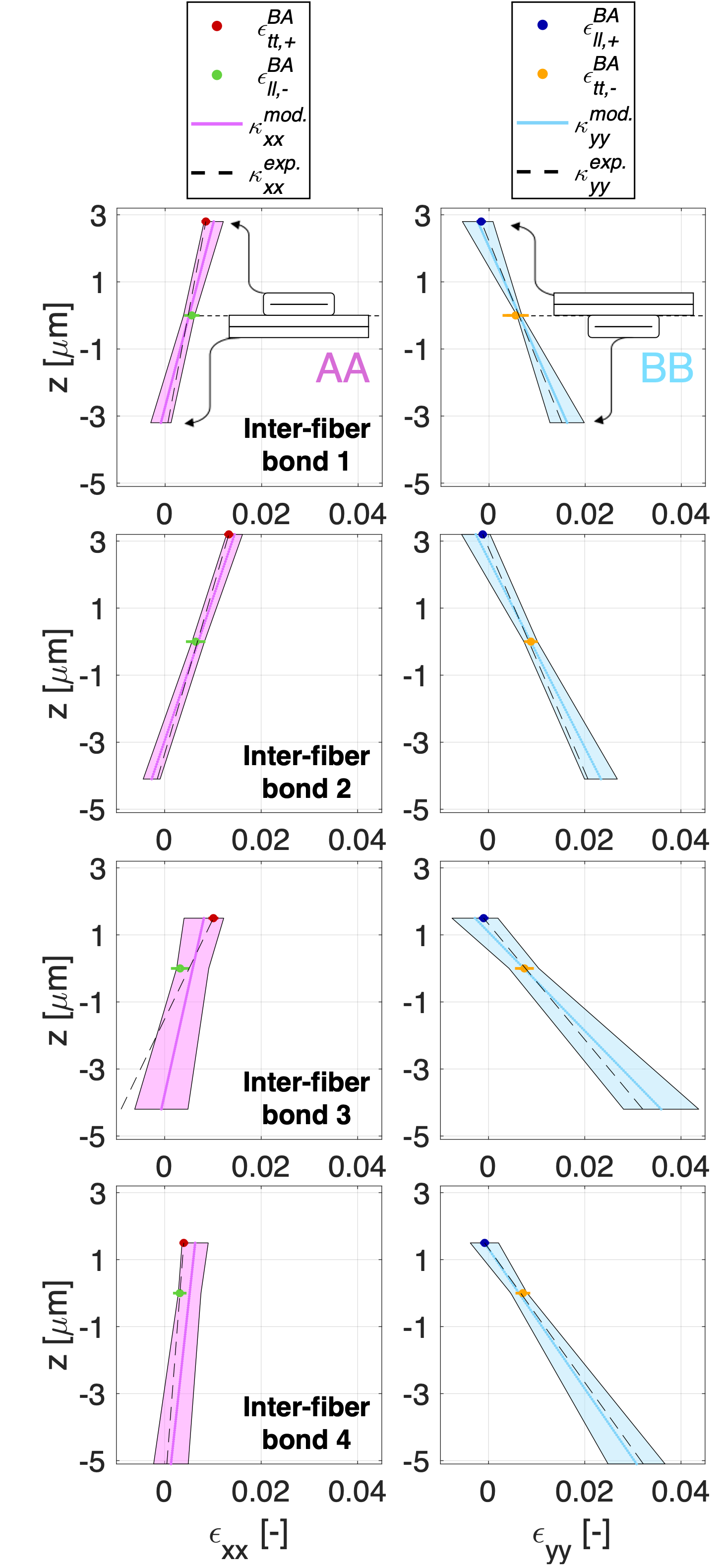}
	\caption{Comparison between the bi-layer laminate model, based on the strains in the free-standing arms of the fibers, (colored solid lines with uncertainty band) and average bond strains ($\epsilon_{ll,+}^{BA}$, $\epsilon_{tt,+}^{BA}$, $\epsilon_{ll,-}^{BA}$, and $\epsilon_{tt,-}^{BA}$), measured directly at the bonded area at the top surface ($z=t_+$) and the fiber-to-fiber interface ($z=0$) given in Figure \ref{fig:strains_all}. The curvature of the bonded area (black dashed line) has been independently determined by fitting the GDHC out-of-plane displacement field of ROI 3a, which reveals a good correspondence with determined bond strains. An adequate match is found between the experimental ($\kappa^{exp.}$) and analytical curvature ($\kappa^{mod.}$) for all inter-fiber bonds, except inter-fiber bond 3, which is explained by the poor bonding area (see Table \ref{tab:character}), whereas perfect bonding is assumed in the laminate model.}
	\label{fig:lam_exp}
\end{figure}

Since the average strains in the bonded area at the surface of the top fiber ($\epsilon_{ll,+}^{BA}$ and $\epsilon_{tt,+}^{BA}$) and the interface ($\epsilon_{ll,-}^{BA}$ and $\epsilon_{tt,-}^{BA}$) are known from the full-field characterization displayed in Figure \ref{fig:strains_all}, a comparison with the proposed bending model is conducted for every inter-fiber bond and, which is presented in Figure \ref{fig:lam_exp}. Moreover, the dashed black line represents the curvature that is determined independently by fitting (with a 2\textsuperscript{nd} order polynomial planar fit) the out-of-plane displacement field resulting from the GDHC analysis of ROI 3a, considering only the central part of the bonded area. This curvature line is plotted starting from the measured strain value at the top surface and adequately matches the experimental data points at the fiber-to-fiber interface for each inter-fiber bond, confirming the reliability of the GDHC analysis. The experimentally and analytically determined curvatures with standard deviation are given in Table \ref{tab:strains}. \\ \indent
Looking at inter-fiber bond 1 and 2, the orthogonal inter-fiber bonds, a good match is found between the experimental curvatures ($\kappa^{exp.}$) and those predicted by the laminate model ($\kappa^{mod.}$). The minor mismatches can be attributed to the assumptions made in the model which do not represent the reality, i.e. perfect bonding between the fibers, see Table \ref{tab:character}, and no forces and moments acting on the inter-fiber bond. The minor increase at the edges of the bonded area in the $\epsilon_{ll,+}$ curves in Figure \ref{fig:strains_all} suggests that while the bond is bending upward, the edges of the bonded area are bending downward due to the fixation of the freestanding segments by the nylon wires to the sample holder. This generates an unavoidable moment at the edges of the bonded area resulting in a larger longitudinal strain. Even though the model does not comply with reality on these points, a good match results. The model also predicts the mechanics of inter-fiber bonds 3 and 4, the non-orthogonal inter-fiber bonds, quite well. However, for inter-fiber bond 3, the horizontal strain ($\epsilon_{xx}$) deviates a bit more from the experiments at the bottom of the bottom fiber. The main reason for this seems to be the significantly lower bonding area compared to the other inter-fiber bonds, as given in Table \ref{tab:character}.
\textbf{\begin{table}[H]
		\centering
		\caption{Inter-fiber bond deformation characteristics with the experimentally ($\kappa^{exp.}$) and analytically ($\kappa^{mod.}$) determined curvature in horizontal and vertical directions, and the bond strain predicted by the bi-layer model ($\bar{\epsilon}$) in the horizontal and vertical direction, with or without bending ($\kappa = 0$).}
		\label{tab:strains}
		\resizebox{1\linewidth}{!}{\begin{tabular}{@{}ccccccccc@{}}
				\toprule
				\begin{tabular}[c]{@{}c@{}}inter-fiber\\ bond \end{tabular}  &  
				\begin{tabular}[c]{@{}c@{}}$\kappa_{xx}^{exp.}$\\ {[}$m^{-1}${]}\end{tabular}  &  \begin{tabular}[c]{@{}c@{}}$\kappa_{xx}^{mod.}$\\ {[}$m^{-1}${]}\end{tabular} &
				\begin{tabular}[c]{@{}c@{}}$\kappa_{yy}^{exp.}$\\ {[}$m^{-1}${]}\end{tabular}  & \begin{tabular}[c]{@{}c@{}}$\kappa_{yy}^{mod.}$\\ {[}$m^{-1}${]}\end{tabular} &
				\begin{tabular}[c]{@{}c@{}}$\bar{\epsilon}_{xx}$ \\ {[}-{]} \end{tabular} & 
				\begin{tabular}[c]{@{}c@{}}$\bar{\epsilon}_{yy}$ \\ {[}-{]} \end{tabular}  & \begin{tabular}[c]{@{}c@{}}$\bar{\epsilon}_{xx}^{\kappa = 0}$ \\ {[}-{]} \end{tabular} & \begin{tabular}[c]{@{}c@{}}$\bar{\epsilon}_{yy}^{\kappa = 0}$ \\ {[}-{]} \end{tabular} \\ \midrule
				1                &  1346$\pm$36    & 1825$\pm$298  & -2759$\pm$40        & -3101$\pm$956   &0.0020$\pm$0.0012  & 0.0020$\pm$0.0005 & 0.0026$\pm$0.0008     & 0.0029$\pm$0.0007      \\
				2                &  2044$\pm$12           & 2354$\pm$111  & -2958$\pm$39    & -3562$\pm$489  & 0.0022$\pm$0.0013 & 0.0032$\pm$0.0014 & 0.0029$\pm$0.0005      & 0.0038$\pm$0.0004       \\
				3                &  3311$\pm$45         & 1547$\pm$520 & -5839$\pm$35       & -6790$\pm$1126   &0.0026$\pm$0.0033 & 0.0023$\pm$0.0031 & 0.0031$\pm$0.0017     & 0.0090$\pm$0.0006       \\
				4                &  542$\pm$11        & 908$\pm$147    & -4958$\pm$39       & -4749$\pm$982    & 0.0032$\pm$0.0023  & 0.0026$\pm$0.0017 & 0.0022$\pm$0.0024 & 0.0057$\pm$0.0003       \\ \bottomrule
		\end{tabular}}
\end{table}}
It can be concluded that the bi-layer laminate model adequately describes the bond mechanics and the strain profiles through the thickness of isolated inter-fiber bonds. Therefore, the next step is to investigate the transmission of the transverse fiber strain in the bonded area for the practically relevant case that the fiber bond is fully confined in a paper fiber network. 

\subsubsection{Transverse strain transferal in the bonded region}
In order to determine the contribution of the transverse fiber strain to the sheet level, it is essential to determine the so-called transverse strain transferal, i.e. the fraction of the transverse strain of one fiber that is transmitted through the bonded fiber in the bonded area. The transverse strain transferal is obtained by first determining the bond strain, predicted by the bi-layer laminate model, in the "longitudinally oriented fiber", i.e. the bottom ($-t_-<z<0$) and top fiber ($0<z<t_+$) for, respectively, cross-sections AA and BB, as depicted in Figure \ref{fig:lam_mod}. This strain directly contributes to the network, in contrast to the bond strain in the "transversely oriented fiber", i.e. the top ($0<z<t_+$) and bottom fiber ($-t_-<z<0$). Because the strain varies linearly through the thickness of the fiber, the horizontal ($\bar{\epsilon}_{xx}$) and vertical bond strain predicted by the bi-layer laminate model ($\bar{\epsilon}_{yy}$) is determined at, respectively $z = -\frac{1}{2}t_-$ and $z = \frac{1}{2}t_+$ with its standard deviation considering cycles 2$-$4, as given in Table \ref{tab:strains}. Note that without the bi-layer laminate model, the horizontal bond strain in the bottom fiber could not be determined, because the back side of the inter-fiber bond was not captured. \\ \indent
It is generally known from 3D paper characterization using X-ray Computed Tomography that, on average, 50\% of the surface of the fibers inside the paper sheet are bonded to neighboring fibers \citep{borodulina2016extracting, urstoger2020microstructure}. Therefore, inside a paper sheet, the inter-fiber bonds are much more constrained and consequently much less bending compared to the isolated inter-fiber bonds characterized in the experiments above. Therefore, the earlier computed bond strains do not represent the reality in a paper sheet. Hence, the proposed analytical model is adapted to reflect the more realistic zero bending boundary conditions, i.e. Equation \ref{eq:prof} becomes $\epsilon = \epsilon_0$, resulting in a constant strain through the thickness of the inter-fiber bond. After solving the balance equations, constant bond strain values are found, which are added to Table \ref{tab:strains} ($\kappa$ = 0). For inter-fiber bond 1 and 2, consisting of almost equally thick fibers, the bond strain is rather similar in both directions. Whereas for inter-fiber bond 3 and 4, both consisting of a thicker bottom (horizontal) fiber, the vertical strain is significantly larger, implying that the transverse strain of the bottom fiber can easily stretch the relatively thinner top fiber in its longitudinal direction.\\ \indent
To retrieve the actual transverse strain transferal in, e.g., the horizontal direction, the longitudinal strain of the bottom fiber needs to be subtracted from the horizontal bond strain ($\bar{\epsilon}_{xx}^{\kappa =0}$) given in Table \ref{tab:strains}, resulting in an additional strain contribution provided by the transverse strain of the top (vertical) fiber. Subsequently dividing the resulting value by the transverse strain of the top fiber results in a factor ($TST_x$), representing the fraction of the transverse strain of the top fiber that is transferred to the bottom fiber in the bonded area. The same procedure can be followed for the vertical direction to obtain $TST_y$, resulting in
\begin{equation}
	TST_x = \frac{ \bar{\epsilon}_{xx}^{\kappa =0} - \epsilon^f_{ll,-}}{\epsilon^f_{tt,+}}\text{, and }TST_y = \frac{ \bar{\epsilon}_{yy}^{\kappa =0} - \epsilon^f_{ll,+}}{\epsilon^f_{tt,-}}.
\end{equation}
The resulting transverse strain transmission factors $TST_x$ and $TST_y$ are listed in Table \ref{tab:trans}. \\ \indent
\textbf{\begin{table}[]
		\centering
		\caption{Transverse stain transferal factor in horizontal and vertical direction, i.e. $TST_x$ and $TST_y$.}
		\label{tab:trans}
		\resizebox{0.40\linewidth}{!}{\begin{tabular}{@{}ccc@{}}
				\toprule
					inter-fiber bond  &  $TST_x$ {[}-{]} &$TST_y${[}-{]}\\ \midrule
				1                &  0.085$\pm$0.084  & 0.097$\pm$0.051   \\
				2                &  0.076$\pm$0.033   & 0.082$\pm$0.027  \\
				3                &  0.038$\pm$0.064   & 0.302$\pm$0.202   \\
				4                &  0.006$\pm$0.160   & 0.184$\pm$0.034  \\ \bottomrule
		\end{tabular}}
\end{table}}
Inter-fiber bonds 1 and 2 remarkably show very similar values, both between horizontal and vertical directions, as well as between the two inter-fiber bonds, which can be attributed to almost similar fiber thickness, the orthogonal inter-fiber bond geometry, and comparable hygro-expansivity given in Table \ref{tab:character} and \ref{tab:hygro}. The bottom fiber is slightly thicker for both inter-fiber bonds, hence the slightly larger transverse strain transferal in vertical direction. Furthermore, for inter-fiber bond 3 and 4, the transverse strain transferal in horizontal direction is small, because the bottom fiber is significantly thicker than the top fiber, hence also the large vertical transverse strain transmission. The relatively large standard deviation found for the horizontal transverse strain transferal of inter-fiber bond 3 and 4 is attributed to the small bond strain (which is almost equal to the longitudinal strain of the bottom fiber) and the relatively large uncertainty on the longitudinal hygro-expansion of the bottom fiber, propagating to a large uncertainty. Additionally, Figure \ref{fig:strains_all} (c) shows that the longitudinal strain of the bottom fiber is almost equal in the bonded area and the freestanding segments, indicating that the thicker bottom fiber is barely influenced by the thinner top fiber; the same holds for inter-fiber bond 4. In contrast, inter-fiber bond 1 and 2, consisting of two almost equally thick fibers, show a relatively small uncertainty in the transverse strain transferal factor due to the larger bond strain values.\\ \indent
Reflecting on the work of \cite{nanko1995mechanisms}, the authors stated that the transverse strain transferal was not a "100\% efficient", whereby it was estimated to be "less than 50\%". The findings in this work show that the transferal factor is actually much smaller, i.e. 8.5$\pm$4.9\% for inter-fiber bond 1 and 2. The discrepancy can be attributed to the difference in testing methods, i.e. the measurement by Nanko and Wu were limited to (i) two extreme situations, i.e. wet and dry during which the inter-fiber bond morphology (bonded area, degree of wrap around) may significantly change, in contrast to the inter-fiber bonds made prior to testing in this work, of which the geometry remains preserved during testing; (ii) single cycle (from wet to dry) testing during which manufacturing errors affect the outcome, whereas we systematically observe that the first cycle always deviates from the successive cycles (e.g., Figure \ref{fig:strains_fiber} and \ref{fig:strains_bend}), which motivates the choice in this work to only include the last three out of four cycles, reducing the influence of the dried-in strain due to manufacturing; (iii) measurement of fibers at the paper surface only, not taking into account possible bending effects which may exist, in contrast to inter-fiber bonds that are representative for a location deeper into the paper microstructure as considered here; and (iv) lack of knowledge of the full geometry of the characterized inter-fiber bonds, in contrast to this work. These limitations may also explain the large spread in the bond strain of 1$-$15\% found by \cite{nanko1995mechanisms}. In contrast to the work of \cite{nanko1995mechanisms}, \cite{brandberg2020role} showed through 3D network modeling that only 4\% of the transverse fiber hygro-expansivity is taken up by the network, implying that the transverse strain transferal is relatively low, similar to the experimental observations presented here. Additionally, in \citep{vonk2023frc, vonk2023res, vonk2023hydro}, in which single fibers isolated from prepared handsheets were tested, revealed that the longitudinal fiber hygro-expansion is only slightly lower than the sheet-scale hygro-expansivity, confirming the relatively low contribution of the transverse hygro-expansion to the sheet found in this work and in the network modeling work by \cite{brandberg2020role}. Finally, note that the laminate model proposed here can be extended to predict the sheet scale hygro-expansion, for instance by means of the orthogonal network model \citep{bosco2015explaining} consisting of two fibers in which the zero bending laminate model is used to predict the bond strain and a representative porosity is adopted.   

\section{Conclusion}
The contribution of the transverse fiber hygro-expansion to the sheet scale has been under debate in the literature for many years. Therefore, in the proposed work, the full-field hygro-expansivity of four isolated inter-fiber bonds was experimentally characterized, allowing identification of the 3D morphological change, transverse strain transferal, and strain gradients in and around the bonded region.\\ \indent
After manufacturing (near) orthogonal inter-fiber bonds, each bond was clamped to a sample holder which allowed double sided imaging of the front and back to determine the fiber thickness and width, degree of wrap around, contact surface and fiber-to-fiber bond angle, which are essential for a proper understanding of the bond hygro-mechanics, as the fiber-to-fiber deviations are rather large. A recently developed Global Digital Height Correlation (GDHC) algorithm dedicated to fiber swelling was extended to determine the full-field hygro-expansion of the inter-fiber bond in and around the bonded area. All inter-fiber bonds revealed clear bending deformation in the bonded area where the longitudinal direction of one fiber is attached to the transverse direction of the other fiber, which is driven by the significantly larger transverse compared to longitudinal hygro-expansivity of the (orthogonally) bonded fibers. Furthermore, the GDHC results showed a transverse strain decrease near the bonded area, which is attributed to the low longitudinal hygro-expansion of the bonded fiber. As the top fiber covers the bottom fiber in the bonded region, the strain field at the fiber-to-fiber interface surface is unknown, but needed for understanding the bond (hygro-)mechanics. Therefore, interpolation of the strain field on the left and right of the bonded area was employed to determine the strain field at the interface surface, revealing an increased longitudinal strain of the bottom fiber, which is attributed to the large transverse strain of the top fiber that stretches the bottom fiber in its longitudinal direction. Finally, all inter-fiber bonds showed similar deformations, highlighting the method's accuracy and the reproducibility of the inter-fiber bond (hygro-)mechanics, even for bonds consisting of significantly deviating fiber dimensions and fiber-to-fiber bond angles. \\ \indent
An analytical bi-layer laminate model was used to describe the bending deformation in the bonded region. This model requires material parameters, dimensions, and the hygro-expansion of free segments of the fibers constituting the bond, all measured in the experiments or known from the literature. This model allows to predict the strain profile in horizontal and vertical direction through the thickness of the bonded area. Because the average bond strain in both directions at two locations through the thickness, i.e., the top fiber surface and the fiber-to-fiber interface surface, of the inter-fiber bond are known from the full-field characterization, a comparison was conducted and an adequate match was found. Moreover, the curvatures predicted by the model are also in good agreement with those obtained directly by fitting the out-of-plane displacement field from GDHC. Hence, the laminate model can predict the bond (hygro-)mechanics well. Different boundary conditions applied to the laminate model allowed to predict the transverse strain transferal, i.e. the fraction of the transverse strain of one fiber stretching the bonded fiber in its longitudinal direction which contributes to the sheet scale hygro-expansion. It was found that, for two inter-fiber bonds of similar fiber dimensions, the transverse strain transferal is relatively weak, which is in close agreement with recent findings from 3D network modeling proposed in the literature.

\end{document}